\documentclass[pra,aps,twocolumn,showpacs]{revtex4-1}
\usepackage{epsfig,amsmath}

\begin{document}

\title{Orbital magnetization of interacting Dirac fermions in graphene}

\author{Xin-Zhong Yan$^{1}$ and C. S. Ting$^2$}
\affiliation{$^{1}$Institute of Physics, Chinese Academy of Sciences, P.O. Box 603, 
Beijing 100190, China\\
$^{2}$Texas Center for Superconductivity, University of Houston, Houston, Texas 77204, USA}

\date{\today}

\begin{abstract}
We present a formalism to calculate the orbital magnetization of interacting Dirac fermions under a magnetic field. In this approach, the divergence difficulty is overcome with a special limit of the derivative of the thermodynamic potential with respect to the magnetic field. The formalism satisfies the particle-hole symmetry of the Dirac fermions system. We apply the formalism to the interacting Dirac fermions in graphene. The charge and spin orderings and the exchange interactions between all the Landau levels are taken into account by the mean-field theory. The results for the orbital magnetization of interacting Dirac fermions are compared with that of noninteracting cases.
\end{abstract}


\maketitle

\section{intrduction}

The study of the properties of interacting Dirac (or Weyl) fermions in (topological) semimetals under a magnetic field is a fundamental subject of the condensed matter physics \cite{Castro,Goerbig}. One of the physical themes is to investigate the orbital magnetization (OM) of the Dirac fermions (DFs) with Coulomb interactions. The OM of an electron system is usually defined as \cite{Shi}
\begin{equation}
M=-(\partial \Omega/\partial B)_{T,\mu}  \label{mgn}
\end{equation}
where $\Omega = \Omega(T,\mu,B)$, as a function of the temperature $T$ and the chemical potential $\mu$ and the magnetic field $B$, is the thermal dynamic potential. Equation (\ref{mgn}) is equivalent to a statistical average of the OM operator \cite{Hirst}. However, for Dirac (or Weyl) fermions, Eq. (\ref{mgn}) is ill defined because the occupation of the Landau levels in the lower band leads to divergence of $\Omega$ and thereby $M$. For noninteracting DFs in graphene, $\Omega$ can be evaluated with a special method \cite{Gusynin,Hesse,SGB,Slizovskiy} by which the field $B$ dependent part of $\Omega$ is separated out. The effects of finite-temperature occupations and the impurity broadening of the Landau levels on the OM of the noninteracting DFs have been studied \cite{SGB,Slizovskiy,Koshino}. Nonetheless, for interacting DFs, it is not easy to separate the $B$-dependent part of $\Omega$ from that of the independent part. Study of the OM of Dirac fermions with Coulomb interactions is lacking. How to calculate the OM of interacting DFs is still an open question. In this paper, we are developing a general approach for solving this problem and use it to calculate the OM of interacting Dirac fermions in graphene. 

\section{formalism}

The electrons in graphene are moving on a honeycomb lattice of carbon atoms. The Hamiltonian of the electrons with a neutralizing background is
\begin{equation}
H=-t\sum_{\langle ij\rangle s}c^{\dagger}_{is}c_{js}+U\sum_{j}\delta n_{j\uparrow}\delta n_{j\downarrow} +\frac{1}{2}\sum_{i\neq j}v_{ij}\delta n_{i}\delta n_{j} \nonumber\\
\end{equation}
where $c^{\dagger}_{is}$ ($c_{is}$) creates (annihilates) an electron of spin $s$ in site $i$, $\langle ij\rangle$ sums over the nearest-neighbor (NN) sites, $t \approx$ 3 eV is the NN hopping energy, $\delta n_{is}=n_{is}-n_s$ is the number deviation of electrons of spin $s$ at site $i$ from the average occupation $n_s$, and $U$ and $v_{ij}$ are the Coulomb interactions between electrons. In real space, $v_{ij} = v(r_{ij})$ with $r_{ij}$ the distance between sites $i$ and $j$ is given by
\begin{equation}
v(r) = \frac{e^2}{r}[1-\exp(-q_0r)], \nonumber\\
\end{equation}
where $q_0$ is a parameter taking into account the wavefunction spreading effect in the short-range interactions between electrons. Here we take $q_0 = 0.5/a_0$ with $a_0 \approx 2.46$ \AA~ as the lattice constant of graphene. For carrier concentration close to the charge neutrality point (CNP), one usually adopts the simplified continuum model. With the continuum model and using the mean-field theory (MFT, or the self-consistent Hartree-Fock approximation), we have recently studied the Landau quantization of the interacting electrons taking into account the charge and spin orderings and the exchange interactions between all the levels \cite{Yan}. 

According to the many-particle theory \cite{Luttinger}, the thermodynamical potential $\Omega$ per unit volume of an electron system under a magnetic field $B$ is given by
\begin{eqnarray}
\Omega &=& k_BT\{\Phi-\frac{B}{2\pi}\sum_{k\omega}\exp(i\omega\eta){\rm Tr}[\Sigma(k,i\omega)G(k,i\omega)  \nonumber\\
&&-\ln(-G(k,i\omega))]\} \label{tmp}
\end{eqnarray}
where $\Phi$ is the `free energy' functional of the Green's function $G$, $\Sigma$ is the self-energy, $k$ is the state index, $\omega$ is the fermionic Matsubara frequency, and $\eta$ is an infinitesimal small positive quantity. For Dirac fermions in graphene, $G$ and $\Sigma$ are $2\times 2$ matrices in the space of sublattices $a$ and $b$, and $k$ stands for $(n,v,s)$ with $n,v,s$ respectively the indexes of the Landau level (LL) and valley and spin \cite{Yan}. The self-energy matrix element $\Sigma_{ll'}(k,i\omega)$ with $l (l') = a$ or $b$ is related with $\Phi$ by
\begin{equation}
\Sigma_{ll'}(k,i\omega) = \delta\Phi/\delta G_{l'l}(k,i\omega), \label{sfe}
\end{equation}
which ensures the microscopic conservation law being satisfied \cite{Baym}. The point here is, after the summation over the Matsubara frequency, $\Omega$ can be expressed as the sum over the LLs from $n = 0$ to $\infty$. We will use the units in which $\hbar = e = c = a_0 = 1$, the energy unit $\epsilon_0 = \hbar v_0/a_0 = 1$ (with $v_0$ the Fermi velocity of electrons in graphene), and the unit of magnetic field $B_0 = \hbar c/ea_0^2 = 1$. 

\begin{figure}[t]
\centerline{\epsfig{file=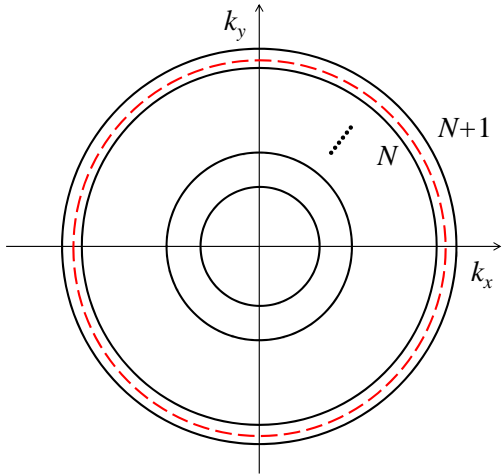,height=8.5cm}}
\caption{(color online) Sketch of Landau levels in momentum space. Under a magnetic field, the states in momentum space are quantized onto the circles. The red dashed circle between the $N$th and $N+1$th Landau levels is the cutoff.} 
\end{figure}

To get rid of the divergence difficulty, we consider a system in momentum space containing finite LLs as shown in Fig. 1. The cutoff momentum is given by $k_c = \sqrt{(2N+1)B}$ where $N$ is the highest Landau index at the field $B$. The thermodynamic potential of this finite system is then given by $\Omega_N(B)$ (suppressing the $T$ and $\mu$ dependence for brevity). The number $N$ changes with $B$ varying for fixed $k_c$. When the magnetic field $B$ varies from $B = k_c^2/(2N+1)$ to $B + \Delta B$ with $\Delta B = 2B/(2N-1)$, the index of the highest LL changes to $N-1$. We then define the OM of the finite system as 
\begin{equation}
M=-\frac{\Omega_{N-1}(B+\Delta B)-\Omega_N(B)}{\Delta B}.    \label {mdf}  
\end{equation}
The ratio given by Eq. (\ref{mdf}) with $k_c \to \infty$ can be considered as the special limit of the derivative in Eq. (\ref{mgn}). For sufficient large cutoff $k_c$, this definition should give rise to the result of the entire system. For low carrier concentration close to the charge neutrality point (CNP), the cutoff can be taken as $k_c = 1$. Here, we should remark that our finite system of $N$ LLs is part of the whole system of infinite LLs. It does not mean we can consider an isolated system of only $N$ LLs from the beginning. For the Dirac fermions, the cutoff for such an isolated system leads to unphysical results. The consideration of such an isolated Dirac system is equivalent to thinking only the top $N$ LLs being occupied with the rest lower LLs as empty in the lower band. This is apparently unphysical. 

Now that $\Omega_N(B)$ contains $N$ terms, we define
\begin{equation}
\Omega_N = BS_N(B)
\end{equation}
and suppose each term in $S_N(B)$ be an analytical function of $B$. Write $S_{N-1}(B+\Delta B) = S_{N}(B+\Delta B)-y_{N}(B+\Delta B)$ with $y_N$ the $N$th term in the sum $S_N$. Then, by expanding $S_{N}(B+\Delta B)$ to order $(\Delta B)^2$ and $y_{N}(B+\Delta B)$ to order $\Delta B$, the OM can be expressed as
\begin{eqnarray}
M &=&-S_N(B)-\frac{2N+1}{2N-1}[BS'_N(B)+\frac{B^2S_N''(B)}{2N-1 }]\nonumber\\
~~&&+(N+1/2)[y_N(B) + By_N'(B)/(N-1/2)]
\label{mgnr}
\end{eqnarray}
where the primes mean the derivatives with respect to $B$. 

\section {OM of noninteracting Dirac fermions}

As an example, here, we consider the free Dirac fermions in graphene at zero temperature. The Hamiltonian of a single Dirac fermion is 
\begin{eqnarray}
H_v(p) = s_vp_x\sigma_1+p_y\sigma_2, \label{hfdf}
\end{eqnarray}
where $s_v = 1$ (-1) for particle in valley $v = K$ ($K'$), the momentum $\vec p$ in each valley is measured from the Dirac point, and $\sigma's$ are the Pauli matrices operating in the sublattice ($a, b$) space. Under a magnetic field $B$ applied perpendicularly to the system plane, the states of the Dirac fermions are given by the Landau quantization. In the LL representation, the Hamiltonian (\ref{hfdf}) reads
\begin{eqnarray}
H_{vn} = \sqrt{2Bn}\sigma_1, \label{ldq}
\end{eqnarray}
where $n$ is the LL index. The LLs are obtained as $\epsilon_{\lambda}(n) = \lambda\sqrt{2Bn}$ with $\lambda = \pm$ for $n \ne 0$, and $\epsilon_0 = 0$ for $n=0$. At CNP and $T = 0$, the LLs in the lower band are fully occupied while the LLs in the upper band are completely empty. The thermodynamic potential reads (see Appendix)
\begin{eqnarray}
\Omega &=& \frac{k_BTB}{2\pi}\sum_{k\omega}\exp(i\omega\eta){\rm Tr}\ln(-G(k,i\omega)) \nonumber\\
&=& \frac{2B}{\pi}\sum_{n}\epsilon_{-}(n), \label{tmpf}
\end{eqnarray}
where the $k$ sum in the first line is understood over the Landau index $n$ and the valley $v$ and the spin $s$. The sum $S_N(B)$ is then obtained as \cite{ram}
\begin{eqnarray}
S_N(B) &=& -c_0\sum_{n=1}^N\sqrt{n} \nonumber\\
 &=& -c_0[\frac{2}{3}(N+1/2)^{3/2}+\zeta(-1/2)+O(\frac{1}{\sqrt{N}})]\nonumber 
\end{eqnarray}
with $c_0 = 2\sqrt{2B}/\pi$ and $\zeta(-1/2) = -0.207886225$. We then have $2BS_N'(B)=-4B^2S_N''(B) = S_N(B)$ and $By_N'(B) = y_N/2 = -c_0\sqrt{N}/2$. 
$M$ is calculated as 
\begin{eqnarray}
M &=&-S_N(B)\{1+\frac{1}{2}\frac{2N+1}{2N-1}[1-\frac{1}{2(2N-1)}]\}\nonumber\\
~~&&-c_0\sqrt{N}(N+1/2)(1 + \frac{1}{2N-1}) \nonumber\\
&=&\frac{3c_0}{2}\zeta(-1/2)+O(\frac{1}{\sqrt{N}}). \label{fdf}
\end{eqnarray}
Due to the expansion of $S_N(B+\Delta B)$ to second order in $\Delta B$ and $y_N(B+\Delta B)$ to linear $\Delta B$, there are precise cancellations from $O(N^{3/2})$ to $O(1)$ in Eq. (\ref{mgnr}). The result given by Eq. (\ref{fdf}) is consistent with the existing one \cite{SGB,Slizovskiy,Ghosal}. 

At finite doping with chemical potential $\mu > 0$ and $T = 0$, the sum $S_N(B)$ is given by
\begin{eqnarray}
S_N(B) = \frac{2}{\pi}[\sum_{n=1}^N(-\sqrt{2Bn}-\mu)+\sum_{n=1}^{N_F}(\sqrt{2Bn}-\mu)-\mu] \nonumber
\end{eqnarray}
where the sums in the square brackets are, respectively, from the lower and upper bands with $N_F$ the index of highest LL below the chemical potential, and the last term $-\mu$ comes from the zero LL. Suppose $\mu << k_c$, we then have $N_F << N$. In the limit $B \to 0$, because of $N_F >> 1$, we obtain 
\begin{eqnarray}
S_N(B) &=& \frac{2c_0}{3}[(N_F+1/2)^{3/2}-(N+1/2)^{3/2}+O(\frac{1}{\sqrt{N_F}})]\nonumber\\
&& -\frac{2}{\pi}(N+N_F+1)\mu. \nonumber 
\end{eqnarray}
The OM is given by  
\begin{eqnarray}
M = \frac{2}{\pi}(N_F+1/2)\{\mu-\sqrt{2B(N_F+1/2)}[1+O(N_F^{-2})]\}. \nonumber 
\end{eqnarray}
As $B \to 0$, $M$ oscillates rapidly between $-\mu/2\pi$ and $\mu/2\pi$ with period $\Delta B = 2B^2/\mu^2$. This is the de Haas-van Alphen oscillation. The average of $M$ vanishes, which leads to the vanishing orbital magnetic susceptibility $\chi = 0$ (defined as the derivative of $M$ with respect to $B$ at $B = 0$) at finite doping. On the other hand, at the CNP, because $M \propto \sqrt{B}$ as given by Eq. (\ref{fdf}), $\chi$ diverges at $B = 0$. This is consistent with the existing result \cite{Koshino,JWM,Safran}, $\chi = -(2/3\pi)\delta(\mu)$, which is obtained by the response of uniform Dirac fermions to the magnetic field without considering the Landau quantization \cite{Safran,Fukuyama}.

\section{MFT for interacting Dirac fermions in graphene}

As in our previous work, we use the MFT to deal with the interactions between the electrons \cite{Yan,Cote}. By the MFT, the `free energy' functional $\Phi$ is approximated as shown in Fig. 2(a). The self-energy is then obtained as in Fig. 2(b), which is independent of the Matsubara frequency. In terms of $G$, $\Phi$ is given by
\begin{eqnarray}
\Phi&=&\frac{B}{4\pi\beta}\sum_{kk',\omega\omega',ll'}e^{i\omega\eta}G_{ll'}(k,i\omega)
v_{l'l}(0)\nonumber\\
&&~~\times e^{i\omega'\eta'}G_{l'l}(k',i\omega') \nonumber\\
&&-\frac{B}{4\pi\beta}\sum_{kk',\omega\omega',ll'}e^{i\omega\eta}G_{ll'}(k,i\omega)
v^x_{l'l}(k,k')\nonumber\\
&&~~\times[e^{i\omega'\eta'}G_{l'l}(k',i\omega')-\beta\delta_{\omega\omega'}\delta_{ll'u}] \nonumber\\
&=&\frac{B\beta}{4\pi}\sum_k{\rm Tr}\{[\Sigma(k)+V^x/2]F(k)\},  \label{phi}
\end{eqnarray}
with $\Sigma(k)=\Sigma_H+\Sigma_X(k)$ and
\begin{eqnarray}
F(k)&=&\frac{1}{\beta}\sum_{\omega}e^{i\omega\eta}G(k,i\omega),\nonumber\\
\Sigma_{H,ll'}&=&\sum_{k'}v_{ll'}(0)F_{ll'}(k') \nonumber\\ 
&=&(v_c\rho_{l}-sUm_{l})\delta_{ll'}, \nonumber\\
\Sigma_{X,ll'}(k)&=&-\sum_{n'}v^x_{ll'}(k,k')[F_{ll'}(k')-\delta_{ll'}/2], \label{sf1}\\
V^x_{ll'}&=&\delta_{ll'}v^x(r)|_{r=0}, \nonumber
\end{eqnarray}
where $\beta = 1/k_BT$, and $\Sigma_{H,ll'}$ has been written in terms of the charge $\rho_{l}$ and the spin $m_{l}$ order parameters with $v_c$ and $U$ the corresponding interaction parameters. The first sum in the first equal of Eq. (\ref{phi}) is due to the direct Coulomb interaction, while the second sum comes from the exchange interaction. Here, $v_{\mu\nu}(0)$ and $v^x_{\mu\nu}(k,k')$ are the interaction elements in the LL representation; they are dependent on the magnetic field $B$ \cite{Yan}. The appearance of the extra term -1/2 in addition to the diagonal distribution function $F_{ll}(k')$ in Eq. (\ref{sf1}) originates from the interaction form of the system of DFs with a neutralizing background given in terms of the density-density multiplication instead of the normal order of the fermion operators. Corresponding to this term, there is a shift $V^x/2$ from the self-energy as shown in Eq. (\ref{phi}); this shift is not drawn in the diagrams in Fig. 2. Because of this shift, the particle-hole symmetry of the system is reflected by the invariance under the transform $\mu \to -\mu$ with $\mu = 0$ at the CNP \cite{Yan3}.

\begin{figure}[t]
\centerline{\epsfig{file=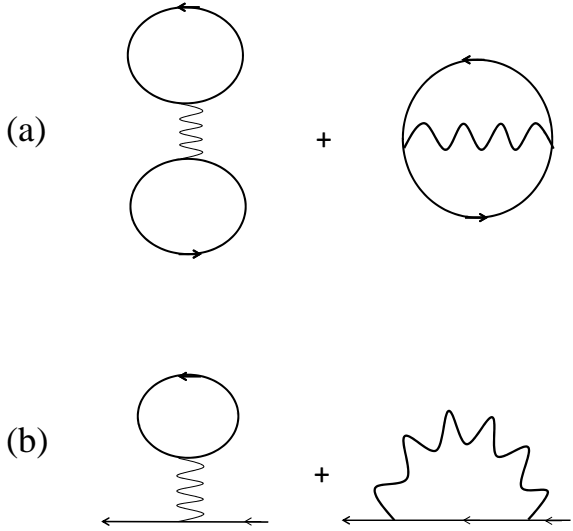,height=8.cm}}
\caption{(color online) (a) `Free energy' functional $\Phi$ under the MFT. (b) Self-energy. The solid line with an arrow denotes the Green's function. The wave line is the interaction. The thick wave line is the exchange interaction including the electron screening effect.} 
\end{figure}

In the LL's picture, the Green's function is given by
\begin{equation}
G(k,i\omega) = \sum_{\lambda}\frac{\psi_{\lambda}(k)\psi^{\dagger}_{\lambda}(k)}{i\omega+\mu-\epsilon_{\lambda}(k)}  \label{grn}
\end{equation}
where $\psi_{\lambda}(k)$ is the $\lambda$th eigen-wave function with eigen-energy $\epsilon_{\lambda}(k)$. The LLs $\epsilon_{\lambda}(k)$ and the wavefunctions $\psi_{\lambda}(k)$ are determined by \cite{Yan}
\begin{equation}
[\sqrt{2Bn}\sigma_1+\Sigma(k)]\psi_{\lambda}(k)=\epsilon_{\lambda}(k)\psi_{\lambda}(k). \label{ll}
\end{equation}
Express the self-energy matrix as $\Sigma(k) = \Sigma_0(k)\sigma_0+\Sigma_1(k)\sigma_1+\Sigma_3(k)\sigma_3$. The energy levels for $n\ne 0$ are obtained as
\begin{eqnarray}
\epsilon_{\lambda}(k)&=&\Sigma_0(k)+\lambda\{[\sqrt{2Bn}+\Sigma_1(k)]^2+\Sigma^2_3(k)\}^{1/2} \nonumber\\
&\equiv&\Sigma_0(k)+\lambda E(k), \label{engy}
\end{eqnarray}
and the corresponding wavefunctions are
\begin{eqnarray}
\psi_{+}(k) &=& 
\left[\begin{array}{c}R_+(k)\\
R_-(k)
\end{array}\right],\nonumber\\
\psi_{-}(k) &=& 
\left[\begin{array}{c}-R_-(k)\\
R_+(k)
\end{array}\right], \label{wvf}
\end{eqnarray}
where $R_{\pm}(k) = \sqrt{1\pm\Sigma_3(k)/E(k)}/\sqrt{2}$. For $n = 0$, the eigenstates are given by
\begin{eqnarray}
\epsilon_0(0Ks)&=&\Sigma_{bb}(0Ks), ~~~~ \psi(0Ks) = 
\left[\begin{array}{c}0\\
1\end{array}\right], \nonumber\\
\epsilon_0(0K's)&=&\Sigma_{aa}(0K's), ~~~~ \psi(0K's) = 
\left[\begin{array}{c}1\\
0\end{array}\right], 
\end{eqnarray}
in valleys $v = K$ and $v = K'$, respectively. The charge and spin orders are calculated by
\begin{eqnarray}
\rho_a&=&\frac{s_0B}{4\pi}\sum_{l\lambda k}s_lf_{\lambda}(k)|\psi_{l\lambda}(k)|^2, \label{rho}\\
m_l&=&\frac{s_0B}{4\pi}\sum_{\lambda k}sf_{\lambda}(k)|\psi_{l\lambda}(k)|^2, \label{ml}
\end{eqnarray}
where $s_0 = \sqrt{3}/2$ is the area of the unit cell, $B/2\pi$ is the spatial degeneracy of the Landau state, $\psi_{l\lambda}(k)$ is the $l$th component of $\psi_{\lambda}(k)$ and $s_l$ = 1 (-1) for $l = a$ ($b$), $s = 1$ (-1) for spin-up (down), and $f_{\lambda}(k) =f(\xi_{\lambda}) = 1/[\exp(\beta\xi_{\lambda})+1]$ with $\xi_{\lambda}(k)=\epsilon_{\lambda}(k)-\mu$ is the Fermi distribution function. 

Here, we need to pay special attention to the equation for the self-energy element $\Sigma_{ab}(k)~[=\Sigma_{ba}(k)]$ or $\Sigma_1(k)$ given by Eq. (\ref{sf1}). Using the wavefunctions given by Eq. (\ref{wvf}), we have
\begin{eqnarray}
F_{ab}(k) = [f_+(k)-f_-(k)]\frac{\epsilon_1(k)+\Sigma_1(k)}{2E(k)},  \label{f1}
\end{eqnarray}
with $\epsilon_1(k) = \sqrt{2Bn}$. Note that $F_{ab}(k)$ goes to $-1/2$ in the limit $n \to \infty$. Equation for $\Sigma_1(k)$ can be written as 
\begin{eqnarray}
\Sigma_1(k) &=& -\sum_{n'\ne 0}v_{ab}^{xv}(n,n')[F_{ab}(k')+1/2]+V^x_1(n)/2,  \nonumber\\
\label{s1}
\end{eqnarray}
with $v_{ab}^{xv}(n,n') = v_{ab}^{x}(k,k')$ and
\begin{eqnarray}
V^x_1(n) &=& \sum_{n'\ne 0}v_{ab}^{xv}(n,n').
\end{eqnarray}
By so doing, the sum over $n'$ in Eq. (\ref{s1}) converges fast. For $\Sigma_{X,ll}(k)$, Eq. (\ref{sf1}) is the proper form since $F_{ll}(k)-1/2$ goes to zero in the limit $n \to \infty$ and therefore the sum over $n'$ converges quickly. Usually, the self-energy given by Eq. (\ref{sf1}) is evaluated with a cutoff $k_c = 1$ \cite{Barlas,Hwang,Kusminskiy}. By the similar treatment, we have solved Eq. (\ref{sf1}) with cutoff $k_c = 1$ for DFs in a magnetic field in our previous work \cite{Yan}. This cutoff has little effect on the low energy levels close to zero. However, it influences substantially the high levels. In particular, the LLs at the cutoff are strongly modified. As indicated in Sec. II, we should solve the equations of the self-energy for the LLs in the whole range $0 \le n < \infty$. Therefore, the revision given by Eq. (\ref{s1}) is necessary. The big task now is to calculate $V^x_1(n)$.

To calculate $V^x_1(n)$, we first consider the case of $B = 0$ and look for an approximation scheme from the result. By the transform $T(\phi_v) = Diag[1,\exp(i\phi_v)]$ with $\phi_v$ the angle of momentum $(s_v k_x,k_y)$, the effective mean-field Hamiltonian reads
\begin{eqnarray}
T^{\dagger}(\phi_v)H^{vs}(\vec k)T(\phi_v) = k\sigma_1 + \Sigma^{vs}(k),  
\end{eqnarray}
which is independent on the angle $\phi_v$. Here, $k$ is understood as the momentum. The self-energy element $\Sigma^{vs}_{ab}(k)$ reads
\begin{eqnarray}
\Sigma^{vs}_{ab}(k) = -\frac{1}{V}\sum_{k'}v^{x}(|\vec k-\vec k'|)\cos\theta F^{vs}_{ab}(k'),  
\label{sk1}
\end{eqnarray}
where $V$ is the volume (area) of the two-dimensional system, $\theta$ is the angle between $\vec k$ and $\vec k'$, and $F^{vs}_{ab}(k)$ has the same form as given by Eq. (\ref{f1}) provided the Landau energy $\epsilon_1(k)$ is replaced with the energy $k$. We can revise Eq. (\ref{sk1}) to get a similar form as Eq. (\ref{s1}) and obtain the corresponding $V_1(k)$ as
\begin{eqnarray}
V_1(k) &=& \frac{1}{V}\sum_{k'}v^{x}(|\vec k-\vec k'|)\cos\theta \nonumber\\
&=&\int^{\infty}_0\frac{dq}{2\pi}v^{x}(q)f(k,q),  \label{vk1}
\end{eqnarray}
where $f(k,q) = [(k-q)K(\alpha)+(k+q)E(\alpha)]/\pi k$ with $K(\alpha)$ and $E(\alpha)$ the elliptic integrals and $\alpha = 2\sqrt{kq}/(k+q)$. Now, for the quantized interaction $V^x_1(n)$, a reasonable approximation is to replace the continuous momentum $k$ with the quantized one $k_n =\sqrt{2Bn}$ in $V_1(k)$,
\begin{eqnarray}
V^x_1(n) \approx V_1(k_n). 
\end{eqnarray}
 
\section{thermodynamic potential}

Using the result (see Appendix)
\begin{eqnarray}
\sum_{\omega}\exp(i\omega\eta){\rm Tr}\ln[-G(k,i\omega)]
&=& \sum_{\lambda}\ln[e^{-\beta\xi_{\lambda}(k)}+1],\nonumber
\end{eqnarray}
we obtain $\Omega(B)$ under the MFT as
\begin{eqnarray}
\Omega(B)&=& -\frac{B}{4\pi}\sum_{k}\{{\rm Tr}[(\Sigma-V^x/2)F(k)] \nonumber\\
&&+\sum_{\lambda}\frac{2}{\beta}\ln[e^{-\beta\xi_{\lambda}(k)}+1]\}. \label{omg}
\end{eqnarray}
From Eq. (\ref{omg}), we may get $\Omega_N(B)$. However, to maintain the particle-hole symmetry in $\Omega_N(B)$, we must revise the form. 

We need to write the equations for the self-energy $\Sigma_{0,3}(k)$ more clearly
\begin{widetext}
\begin{eqnarray}
\Sigma_0(k) &=& -sUm_0-{\sum_{n'}}'\{[v_{11}^v(n,n')+v_{22}^v(n,n')][g_+(k')+g_-(k')]/4 \nonumber\\
&&+[v_{11}^v(n,n')-v_{22}^v(n,n')][g_+(k')-g_-(k')]\Sigma_3(k')/4E(k')\}-g_0(0vs)v^K_{22}(n,0)/2,  \nonumber\\
\Sigma_3(k) &=& v_c\rho-sUm_3-{\sum_{n'}}'\{[v_{11}^v(n,n')-v_{22}^v(n,n')][g_+(k')+g_-(k')]/4 \nonumber\\
&&+[v_{11}^v(n,n')+v_{22}^v(n,n')][g_+(k')-g_-(k')]\Sigma_3(k')/4E(k')\}+s_vg_0(0vs)v^K_{22}(n,0)/2,  \label{s0}
\end{eqnarray}
where $\rho = \rho_a, m_{0,3} = (m_a \pm m_b)/2, g_{\lambda}(k)=f_{\lambda}(k)-1/2$. Note that $v^K(n,n') = \sigma_1v^{K'}(n,n')\sigma_1$, we then obtain
\begin{eqnarray}
{\sum_k}'\Sigma_0(k)+\sum_{vs}[\Sigma_0(0vs)-s_v\Sigma_3(0vs)]/2 = -V^x{\sum_{k\lambda}}'g_{\lambda}(k)/2-V^x\sum_{vs}g_0(0vs)/2, \label{rlt}
\end{eqnarray}
where we have used the relation 
\begin{eqnarray}
\sum_{n'}v^K_{bb}(n,n') = V^x
\end{eqnarray}
which is independent on $n$. Using Eq. (\ref{rlt}), we rewrite Eq. (\ref{omg}) in the form
\begin{eqnarray}
\Omega(B) &=&-\frac{B}{2\pi}{\sum_{k}}'\{\sum_{\lambda}\frac{1}{\beta}\ln(e^{\beta\xi_{\lambda}/2}+e^{-\beta\xi_{\lambda}/2})
+\Sigma_0(F_0-1/2)+\Sigma_1F_1+\Sigma_3F_3+\mu-V^x/4\}\nonumber\\
&& -\frac{B}{2\pi}\sum_{vs}\{\frac{1}{\beta}\ln(e^{\beta\xi_0/2}+e^{-\beta\xi_0/2})
+(\Sigma_0-s_v\Sigma_3)(f_0-1/2)/2+\mu/2-V^x/8\}, \label{omg1}
\end{eqnarray}
\end{widetext}
where $F_{0,1,3}$ are distribution functions defined as
\begin{eqnarray}
F_0&=&[f_+(k)+f_-(k)]/2,  \nonumber\\
F_1&=& \frac{\epsilon_0(k)+\Sigma_1(k)}{E(k)}[f_+(k)-f_-(k)]/2,  \nonumber\\
F_3&=& \frac{\Sigma_3(k)}{E(k)}[f_+(k)-f_-(k)]/2,  \nonumber 
\end{eqnarray}
$f_0 = f_0(0vs)$ is the Fermi distribution function of level $n=0$, and ${\sum_k}'$ means $n\ne 0$. 

Under the transform $\mu \to -\mu$, the self-energy components change as $\Sigma_{0}(nvs) = -\Sigma_{0}(nvs)$ and $\Sigma_1(nvs) = \Sigma_1(nvs)$ and $\Sigma_3(nvs) = -\Sigma_3(n\bar vs)$ or $\Sigma_{1,3}(nvs) = \Sigma_{1,3}(n\bar vs)$ (with $\bar v$ means $\bar K = K'$ and $\bar K' = K$). Note that the constant terms $\mu-V^x/4$ and $\mu/2-V^x/8$ in Eq. (\ref{omg1}) will disappear in the final formula for $M$ because a cancellation between the terms $-S_N(B)$ and $(N+1/2)y_N(B)$ as indicated by Eq. (\ref{mgnr}). We can then conclude that $M$ is symmetric under the particle-hole transform. The function $S_N(B)$ can now be extracted from Eq. (\ref{omg1}). 

For calculating $S_N'(B)$ and $S_N''(B)$, we need to derive the equations of the self-energy elements with respect to $B$ and solve them. The derivation is elementary but tedious. For brevity of the paper, we will not express these equations here.

\section{orbital magnetization}

We have numerically solved the equations for the self-energy $\Sigma(k), \partial\Sigma(k)/\partial B$, and $\partial^2\Sigma(k)/\partial B^2$. In the present calculation, the on-site interaction is set as $U/\epsilon_0 = 2$. The coupling constant of the interaction is $e^2/a_0\epsilon_0 = 2.2$. With the results for the self-energy and its derivatives, we calculate the OM at CNP and at finite carrier concentration.

Shown in Fig. 3 are the numerical results for the interacting and free DFs at CNP and at $T =0$. It is seen that the magnitude of the OM of interacting DFs (blue solid circles) is smaller than that of the free DFs (red circles). At $T = 0$, there exists antiferromagnetic spin ordering in the interacting DFs catalyzed by the magnetic field as investigated in many works \cite{Yan,Gusynin1,Herbut,Kharitonov,Lado,Roy,Khveshchenko,Alicea,Gusynin2,Jung,Lukose}. This spin ordering results in the splitting of the zero Landau levels. In the low energy zero-LL states, the spin-up and down electrons move in the sublattices $a$ and $b$, respectively. The spin ordering also modifies the electron distributions in the two sublattices at other LLs. Overall, in the presence of the spin ordering, the electrons cannot move freely in the whole lattice. Since the antiferromagnetic spin ordering acts as the obstacle for the orbital circumnutation, the OM is therefore weakened. 

\begin{figure}[t]
\centerline{\epsfig{file=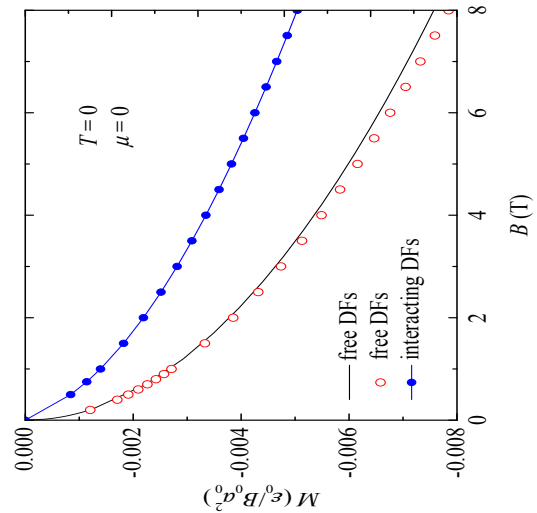,width=8.cm,height=8.cm,angle=0}}
\caption{(color online) Orbital magnetization of interacting Dirac fermions (blue solid circles with line) compared with the result for free Dirac fermions (red circles) at CNP and at $T=0$. The black line represents the analytical result given by Eq. (\ref{fdf}) with $N \to \infty$ for the free Dirac fermions.} 
\end{figure}

The black line in Fig. 3 represents the analytical formula Eq. (\ref{fdf}) for $N \to \infty$. In the numerical calculation, the cutoff $N$ is finite given by $N = k_c^2/2B - 1/2$ with $k_c = 1$ (and $B$ in units of $B_0 = 1.1\times 10^4$ T). At small $B$ close to zero, since $N$ is sufficiently large, the numerical result (red circles) for the free DFs is in very good agreement with the analytical formula. The difference between them increases with increasing $B$. For $B \sim 8$ T, the numerical result seems still good.

In Fig. 4, we present the results at $T/\epsilon_0 = 0.01$ and at CNP. Since $T$ is high, there are many LLs within the temperature range. As a result, the OM of free DFs varies linearly with $B$ consistent with the existing result \cite{Slizovskiy}. While for the interacting DFs, the OM is not linear in $B$ and its magnitude is larger than that of free DFs. At this high temperature, the spin ordering vanishes but the LLs of the DFs are strongly changed by the interactions through the self-energy $\Sigma_1$. $\Sigma_1$ gives rise to an enhancement of the velocity \cite{Yan2,Sarma,Menezes}, leading to fast orbital circumnutations. The nonlinear behavior of $M$ with $B$ implies the renormalized velocity varies with momentum. Because of the vanishing of spin ordering and the enhancement of the velocity, the OM of the interacting DFs is stronger than that of the free DFs.  

\begin{figure}[t]
\centerline{\epsfig{file=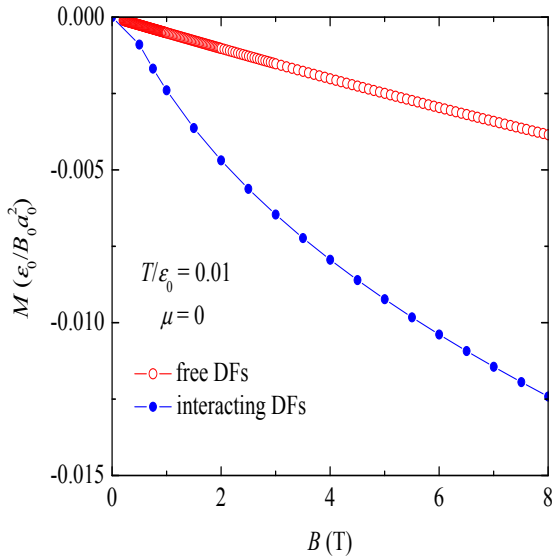,width=8.5cm,height=8.5cm,angle=-90}}
\caption{(color online) Orbital magnetization of interacting Dirac fermions (blue solid circles with line) compared with the result for free Dirac fermions (red circles) at CNP and at $T=0.01$.} 
\end{figure}

Figure 5 exhibits the result for the OM of the interacting DFs at finite carrier concentration with $\mu/\epsilon_0 = 0.02$ and at $T/\epsilon_0 = 0.001$. The chemical potential for the free DFs is set as $\mu_0/\epsilon_0 =0.00167$ so that the first LL in the upper band for both interacting DFs and the free DFs has almost the same position $B$. At the finite carrier concentration and temperature, the charge and spin orderings disappear and all the LLs are degenerated with degeneracy 4. The index of the first LL in the upper band is $n = 1$. With the field $B$ varying, when the LLs pass cross the Fermi level, the OM shows the de Haas-van Alphen oscillations. The LL of $n = 1$ is at about $B \approx 0.97$ T. Above this field, there are no LLs below the Fermi level in the upper band and the OM decreases monotonically with $B$.

For finite doping at very small $B$ and $T = 0$, there are rapid de Haas-van Alphen oscillations similarly as that indicated in Sec. III for noninteracting DFs. For interacting DFs, however, the average of the oscillations should not be vanishing at small $B$. According to the perturbation theory, the system shows orbital paramagnetism at very small $B$ \cite{Principi}; with the first order perturbation calculation for Thomas-Fermi screened Coulomb interactions, it has been shown that the orbital magnetic susceptibility $\chi$ is positive for DFs in doped graphene. Therefore, the average $M$ should increase from $M = 0$ with increasing the field $B$. At finite $T$, the oscillations are smeared by temperature. The average $M$ should be weakened by the thermal fluctuations. (We did not perform the calculation at very small $B$ because for which the cutoff number $N$ is so large that the accuracy requirement for the numerical calculation exceeds the ability of our computer.)  

\begin{figure}[t]
\centerline{\epsfig{file=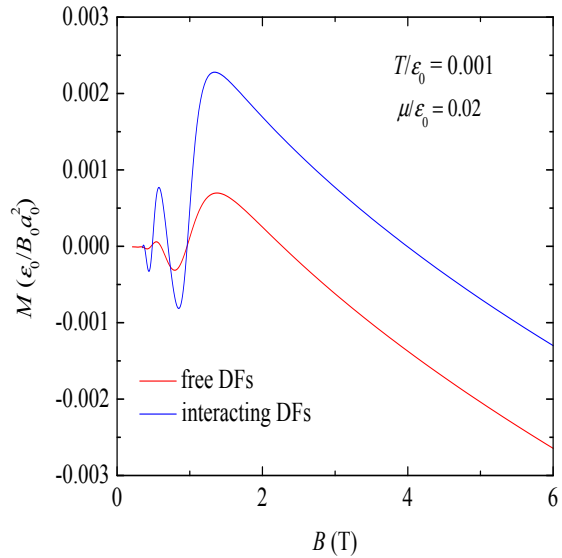,width=8.5cm,height=8.5cm,angle=-90}}
\caption{(color online) Orbital magnetization of interacting Dirac fermions (blue solid line) at $T/\epsilon_0=0.001$ and $\mu/\epsilon_0 = 0.02$compared with the result for free Dirac fermions (red line) with $\mu_0/\epsilon_0 = 0.00167$. } 
\end{figure}

\section{Remark}

In the present approach, the OM is calculated by expanding the sum $S_N(B+\Delta B)$ [and the term $y_N(B+\Delta B)$] to second (first) order in $\Delta B$ as shown in Eq. (\ref{mgnr}). The formalism works only for the system of eigen-energy being linear in momentum $k$. For a Dirac or Weyl system of $\epsilon_{\lambda}(k) \to \lambda k^{\nu}$ as $k \to \infty$, the sum $S_N$ is order $N^{\nu/2+1}$. We need to expand $S_N(B+\Delta B)$ [$y_N(B+\Delta B)$] to $m$th ($m-1$th) order in $\Delta B$ with $m = [\nu/2]+2$. Here $[\nu/2]$ means the integer part of the number $\nu/2$. For example, for an $L$-layered graphene, since it has $[L/2]$ bilayer bands and $L$ mod 2 monolayer bands \cite{Koshino1}, we need to expand $S_N(B+\Delta B)$ to $(\Delta B)^3$ and $y_N(B+\Delta B)$ to $(\Delta B)^2$. By so doing, the unphysical part will be eliminated due to the precise cancellations between these expanded terms.

Though the system of infinitive LLs is considered, the contribution to the total OM comes mostly from the LLs below $k_c = 1$ as reflected by the result for free DFs shown in Fig. 3. In graphene, the Dirac cone approximation to the energy bands of electrons is valid within the circle of radius $k_c=1$ in the momentum space. Therefore, the present result for the OM is a fairly good measure of that of electrons in graphene. However, as already stressed, we cannot isolate the LLs below $k_c$ from the entire system. The reason is that the high LLs (especially the LLs close to the cutoff) are strongly modified by the isolation. We have performed the numerical calculation for the isolated system. The consequence of the isolated system is that the magnitude of the OM is several orders larger than the result presented here; it becomes bigger and bigger as $B \to 0$ even not vanishing at $B = 0$. 

\section{conclusion}

We have developed the approach for calculating the orbital magnetization of Dirac fermions. The main points in the formalism are: (1) To overcome the divergence difficulty due to the occupation in the lower band, the orbital magnetization is defined as the special limit for the derivative of the thermodynamic potential with respect to the magnetic field. (2) The equations for the self-energy and its derivatives with respect to the magnetic field need to be solved. (3) The particle-hole symmetry should be ensured in the partly sum of the thermodynamic potential. (4) The system with finite LLs is part of the entire system but not isolated from the rest of the entire system. 

With the formalism, we have calculated the OM for interacting DFs in graphene and compared the results with that of the free DFs. At very low carrier concentration close to CNP, when the antiferromagnetic spin ordering catalyzed by the magnetic field exists, the OM is weakened. Without the spin and charge orderings, the OM is enhanced due to the velocity renormalization by interactions. At low temperature and finite carrier concentration, the de Haas-van Alphen oscillation appears in the OM as a function of magnetic field.

The present approach may be extended to study the OM of Weyl fermions in the topological semimetals as well.

\acknowledgments

This work was supported by the National Basic Research 973 Program of China under Grant No. 2016YFA0202300 and the Robert A. Welch Foundation under Grant No. E-1146.

\appendix
\section{}
\renewcommand{\theequation}{\thesection\arabic{equation}}
\setcounter{equation}{0}

{\it 1. Matsubara-frequency sum.} For calculating $\Omega$, we need the following sum,
\begin{eqnarray}
\frac{1}{\beta}\sum_{\omega} \exp(i\omega\eta)\ln[-G(k,i\omega)],\nonumber
\end{eqnarray}
which is usually performed with the loop integral in the complex $z$ plane,
\begin{eqnarray}
&& -\oint\frac{dz}{2\pi i}\frac{e^{z\eta}}{e^{\beta z}+1}\ln[-G(k,z)]\nonumber\\
&=& -\int_{-\infty}^{\infty}\frac{d\omega}{\pi}f(\omega)e^{\omega\eta}{\rm Im}\ln[-G(k,\omega^+)]\nonumber\\
&=& -\sum_{\lambda}\psi_{\lambda}(k)\psi^{\dagger}_{\lambda}(k) \int_{-\infty}^{\infty}\frac{d\omega}{\pi }f(\omega)e^{\omega\eta}{\rm Im}\ln[\frac{1}{\xi_{\lambda}(k)-\omega^+}] \nonumber\\
&=& -\sum_{\lambda}\psi_{\lambda}(k)\psi^{\dagger}_{\lambda}(k) \ln\{\exp[-\beta\xi_{\lambda}(k)]+1\}/\beta\nonumber
\end{eqnarray}
where $f(\omega) = 1/[\exp(\beta\omega)+1]$ and $\omega^+ =\omega+0^+$. At $T = 0$, it reduces to
\begin{eqnarray}
\Rightarrow \sum_{\lambda}\psi_{\lambda}(k)\psi^{\dagger}_{\lambda}(k) [\epsilon_{\lambda}(k)-\mu]|_{\epsilon_{\lambda}(k)<\mu}. \label{logg}
\end{eqnarray}

{\it 2. Exchange interaction.} The exchange interaction should contain the screening effect due to the electron density fluctuations. For qualitatively reflecting the screening, we adopt the Thomas-Fermi form for exchange interaction given as
\begin{eqnarray}
v^{x}(q) &=& \frac{v(q)}{1+q_{TF}/q} \label{vsc}
\end{eqnarray}
where $v(q)$ is the Fourier transform of the interaction $v(r)$. In the continuum model, $v(q)$ is given by
\begin{eqnarray}
v(q) &=& \frac{2\pi e^2}{q}-\frac{2\pi e^2}{\sqrt{q^2+q^2_0}}. \nonumber
\end{eqnarray}
By using the long-wavelength limit of the density-density response function of free Dirac fermions with chemical potential $\mu$ at temperature $T$ \cite{Yan1}, the Thomas-Fermi wave number $q_{TF}$ is obtained as
\begin{eqnarray}
q_{TF} &=& \frac{4 e^2}{v^2_0}[|\mu|+2T\ln(1+e^{-|\mu|/T})]. \nonumber
\end{eqnarray}

\end{document}